\documentclass[amsmat,amssymb,amsfonts,aps,prb,twocolumn,showpacs]{revtex4}

\usepackage{graphicx}
\usepackage{dcolumn}
\usepackage{bm}

\begin{document}

\newcommand{\bra}[1]{\langle#1|}
\newcommand{\ket}[1]{|#1\rangle}
\newcommand{\sbra}[1]{\left\langle#1\right|}
\newcommand{\sket}[1]{\left|#1\right\rangle}
\newcommand{\bracket}[2]{\big\langle#1 \bigm| #2\big\rangle}

\title{
  Orbital selective and tunable Kondo effect of magnetic adatoms on graphene:\\
  Correlated electronic structure calculations
}

\author{ D. Jacob } 
\affiliation{Max-Planck-Institut f\"ur Mikrostrukturphysik, Weinberg 2, 06120 Halle, Germany} 
\author{G. Kotliar }
\affiliation{Dept. of Physics \& Astronomy, Rutgers University, 136 Frelinghuysen Rd., Piscataway, NJ 08854, USA}

\date{\today}

\begin{abstract}
  We have studied the effect of dynamical correlations on the electronic structure 
  of single Co adatoms on graphene monolayers with a recently developed novel method
  for nanoscopic materials that combines density functional calculations with a fully 
  dynamical treatment of the strongly interacting $3d$-electrons. 
  The coupling of the Co $3d$-shell to the graphene substrate and hence the dynamic 
  correlations are strongly dependent on the orbital symmetry and the system parameters
  (temperature, distance of the adatom from the graphene sheet, gate voltage).
  When the Kondo effect takes place, we find that the dynamical correlations give
  rise to strongly temperature-dependent peaks in the Co $3d$-spectra near the Fermi 
  level. Moreover, we find that the Kondo effect can be tuned by the application 
  of a gate voltage. It turns out that the position of the Kondo peaks is pinned to 
  the Dirac points of graphene rather than to the chemical potential. 
\end{abstract}

\maketitle

%%%%%%%%%%%%%%%%%%%%%%%%
\section{Introduction}
\label{sec:introduction}
%%%%%%%%%%%%%%%%%%%%%%%%

Since its recent discovery, Graphene ---a monolayer of graphite--- has 
become the subject of intense research due to its peculiar electronic properties 
\cite{Novoselov:science:2004,Novoselov:nature:2005,Zhang:nature:2005,Novoselov:natphys:2006}.
The unusual properties of graphene are largely a result of the linear dispersion 
of electron bands at low energies and its two-dimensionality. This allows to describe
the charge carriers in graphene as a two-dimensional gas of massless Dirac-fermions. 
The unusual electronic properties and the fact that its chemical potential can easily 
be tuned by a gate voltage make graphene an ideal experimental testing ground for exotic 
physics as well as a promising basis for novel nano-electronics devices.
For recent reviews of this rapidly developing field and additional references,
see e.g. Refs. \onlinecite{Beenakker:rmp:2008,Castro-Neto:rmp:2009,Vozmediano:2010}.

Graphene is also a good testing ground for exotic Kondo physics\cite{Hewson:book}:
First, the linearly vanishing DOS of graphene favours the formation of a local 
moment \cite{Uchoa:prl:2008} which is a necessary condition for Kondo physics. 
Also vacancies and edge defects in graphene have been predicted to give rise to 
magnetic moments \cite{Lehtinen:prl:2004,Palacios:prb:2008}. Very recently, 
the Kondo effect has actually been observed at local defects in graphene providing evidence 
that defects in graphene are indeed magnetic \cite{Chen:2010}.
On the other hand, as a result of the linearly vanishing DOS, a magnetic moment 
coupled to a graphene layer can only display a Kondo effect if the strength of 
the coupling is strong enough\cite{Withoff:prl:1990,Hentschel:prb:2007,Sengupta:prb:2008}. 
Furthermore because of the existence of two inequivalent Dirac cones, it has multiple 
channels which can in principle screen the impurity spin and could therefore lead to a 
multichannel Kondo effect and hence non-fermi liquid behaviour
\cite{Sacramento:prb:1991,Sacramento:advphys:1993,Cox:advphys:1998,Cox:jphyscm:1996,Varma:physrep:2002}.
Kondo effect and the possibilty of multichannel Kondo effect in graphene 
have been discussed in a number of recent publications
\cite{Cornaglia:prl:2009,Zhu:2009,DellAnna:jstat:2010,Wehling:prb:2010a,Wehling:prb:2010b,Vojta:2010}.

Moreover, it is possible to probe the Kondo effect of magnetic adatoms 
on graphene by scanning tunneling microscopy (STM). Similar to STM spectroscopy 
of magnetic adatoms and molecules on metallic surfaces
\cite{Madhavan:science:1998,Neel:prl:2007,Vitali:prl:2008,Neel:prl:2008}, 
the Kondo effect could show up as Fano lineshapes in the conductance vs. 
bias voltage characteristics of the STM or as additional temperature dependent peaks in the  STM spectra. 
But in comparison with normal metallic surfaces the Fano lineshapes for 
graphene are expected to be more asymmetric and strongly dependend on 
the symmetry of the $d$-orbitals of magnetic atom \cite{Wehling:prb:2010b}.

The goal of this work is to perform a realistic study of the Kondo effect for a single Co adatom on 
an otherwise perfect graphene sheet by using a recently developed approach for nanoscopic systems 
that combines {\it ab initio} electronic structure calculations on the level of density functional
theory (DFT) calculations with a full dynamical treatment of the strongly interacting $3d$-electrons
\cite{Jacob:prl:2009}.
A realistic treatment allows us to investigate under which conditions the Kondo effect takes place 
in a given orbital and the detailed spectral functions that arise from the Kondo effect in a 
semi-metallic substrate. 

This paper is organized as follows:
In Sec. \ref{sec:method} we briefly describe our recently developed GGA+OCA method\cite{Jacob:prl:2009}
for the quasi {\it ab initio} description of the electronic structure of nanoscopic systems 
fully taking into account dynamical correlations by combining DFT calculations with a dynamical 
treatment of the strongly interacting $3d$-electrons in the so-called one-crossing approximation (OCA).

In Sec. \ref{sec:dft-results} we present DFT electronic structure calculations in the 
generalized gradient approximation (GGA) for determining the most favourable adsorption 
site for the Co atom on the graphene sheet. In agreement with previous results by other 
groups\cite{Mao:jphyscm:2008,Wehling:prb:2010a,Wehling:prb:2010b} we find that the hollow site 
is the most favourable adsorption site. 
Using hybrid functional calculations which generally predict
better geometries for molecules we have then optimized the adsorption height of the Co atom at the 
hollow site of the graphene.
Using GGA we have also calculated the charge and spin of the Co atom in dependence 
of the distance to the graphene sheet for the three different adsorption sites.
The plots indicate that depending on the distance the Co atom can be either in the 
local moment regime or in weak correlation regime. Since the potential energy curves 
are very shallow it seems feasible that the correlation regime be controlled via the 
distance of the Co atom to the graphene substrate in an actual experiment.

In Sec. \ref{sec:oca-results} we study the impact of dynamical correlations of the Co 
$3d$-electrons on the elctronic structure of the Co atom and the graphene sheet using the 
GGA+OCA method for nanoscopic systems as described in Sec. \ref{sec:method}.
We find that the dynamical correlations of the Co $3d$-electrons give rise to strongly
temperature-dependent peaks near the Fermi level in the spectral density of the Co $3d$-shell 
but also in the spectral density of the $p_z$-orbitals of nearby carbon atoms. 
Furthermore, we find that the Co $3d$-spectra depend quite strongly on the exact energies
of the Co $3d$-levels. Accordingly, the application of a gate voltage which 
changes the chemical potential of the graphene substrate has a similar effect on the spectra.
Hence the dynamic correlations and the Kondo effect can be controlled by the application
of a gate voltage.

Finally, in Sec. \ref{sec:conclusions} we conclude this work with a general discussion of the 
results in the light of Kondo physics.

%%%%%%%%%%%%%%%%%%%%%%%%%%%%%%%%%%%%%
\section{GGA+OCA method}
\label{sec:method}
%%%%%%%%%%%%%%%%%%%%%%%%%%%%%%%%%%%%%

We consider a {\it single} Co adatom adsorbed at the hollow site of an otherwise 
perfect graphene sheet as shown in Fig. \ref{fig:graphene+co}.
To proceed we divide the system into two parts. The central region C which contains 
the Co adatom and a number of carbon atoms is embeded into the host material H given 
by a perfect graphene sheet. The central region C has to be chosen big enough 
so that the electronic structure outside of the region can be assumed to be that of 
a perfect graphene sheet. 

As a first step a mean-field description of the electronic structure
of the system is obtained from DFT calculations on the level of GGA using the CRYSTAL06 
{\it ab initio} electronic structure code\cite{CRYSTAL06} together with the 6-31G 
gaussian basis set: By setting up a 2D lattice of supercells of the C region we obtain 
an effective one-body Hamiltonian $\mathbf{H}_{\rm C}$ for the C region from the Kohn-Sham
Hamiltonian of the converged GGA calculation. Analogously, we obtain an effective mean-field 
description of the host material from a GGA calculation for a supercell of clean graphene 
corresponding to the C region.

The Green's function of the central unit cell C is then given by: 
\begin{equation}
  \label{eq:GC}
  \mathbf{G}_{\rm C}(\omega) = (\omega+\mu-\mathbf{H}_{\rm C}+\mathbf{h}_{\rm dcc}-\Sigma_{3d}(\omega)-\mathbf\Sigma_{\rm H}(\omega))^{-1}
\end{equation}
$\mathbf\Sigma_{\rm H}(\omega)$ is the {\it embedding self-energy} which describes the dynamic 
hybridization between the central region C and the host material H. The embedding self-energy 
is obtained from the GGA calculation of the perfect graphene sheet as explained in App. 
\ref{app:embedding}. 

$\Sigma_{3d}(\omega)$ is the electronic self-energy describing the dynamic correlations of the 
strongly interacting $3d$-electrons. The strong interactions of the Co $3d$-shell are captured 
by adding a Hubbard-like interaction term to the one-body Hamiltonian within the correlated 
$3d$-subspace:
\begin{equation}
  \label{eq:HU}
  \hat{\mathcal H}_U = \frac{1}{2} \sum_{{ijkl}\atop{\sigma_1\sigma_2}} U_{ijkl} \,
  \hat{d}_{i\sigma_1}^\dagger\hat{d}_{j\sigma_2}^\dagger \hat{d}_{l\sigma_2}\hat{d}_{k\sigma_1}
\end{equation}
$U_{ijkl}$ are the matrix elements of the effective Coulomb interaction of
the $3d$-electrons which is different from the bare Coulomb interaction due
to the screening by the conduction electrons. Here we take a model interaction
taking into account only the direct Coulomb repulsion $U$ between electrons and 
a Hund's rule coupling $J$. For $3d$ transition metal elements in bulk materials 
the repulsion $U$ is around 2-3~eV and $J$ is around 0.9~eV \cite{Grechnev:prb:2007}.
Due to the lower coordination of the Co atom on the graphene sheet the screening 
should be reduced compared to its bulk value. Here we take $U=5$~eV and $J=0.9$~eV
as in our previous work\cite{Jacob:prl:2009}.

The Coulomb interaction within the correlated $3d$ subspace has already been taken
into account on a static mean-field level in the effective KS Hamiltonian of the
central unit cell C. Therefore the KS Hamiltonian within the correlated subspace 
$\mathbf{h}_{3d}^{\rm KS}$ has to be corrected by a double-counting correction term,
i.e. $\mathbf{h}_{3d}\equiv\mathbf{h}_{3d}^{\rm KS}-\mathbf{h}_{\rm dcc}$.
Here we use the standard expression,
\begin{equation}
  \label{eq:hdcc}
  \mathbf{h}_{\rm dcc} = [U(N_{3d}-1/2) - J(N_{3d}-1)/2]\;\times\;\mathbf{1}_{3d}
\end{equation}
where $\mathbf{1}_{3d}$ is the identity matrix in the Co $3d$-subspace, and $N_{3d}$ is the
occupation of the impurity $3d$-shell \cite{Petukhov:prb:2003}.

In order to compute the electronic self-energy $\Sigma_{3d}$ we have to solve 
the generalized Anderson impurity problem given by the strongly interacting Co 
$3d$-electrons. The impurity problem is completely determined by the interacting 
Hamiltonian of the Co $3d$-shell, i.e. $\hat{\mathcal H}_{3d} = \hat{h}_{3d} + \hat{\mathcal{H}}_U$,
and  by the so-called hybridization function $\mathbf\Delta_{3d}(\omega)$ which 
describes the dynamic hybridization of the Co $3d$-shell with the rest of the system
(i.e. the bath). The hybridization function can be extracted from the non-interacting 
Green's function 
$[\mathbf{G}_{\rm C}^0]^{-1} \equiv [\mathbf{G}_{\rm C}]^{-1}+\mathbf\Sigma_{3d}(\omega)$ 
of the central region C as follows:
\begin{equation}
  \label{eq:Delta}
  \mathbf\Delta_{3d}(\omega) = \omega+\mu-\mathbf{h}_{3d}- [\mathbf{g}_{3d}^0(\omega)]^{-1}
\end{equation}
$\mathbf{g}_{3d}^0$ is the non-interacting Green's function projected onto the $3d$-shell of 
the Co atom, i.e. $\mathbf{g}_{3d}^0\equiv \mathbf{P}_{3d}\mathbf{G}_{\rm C}^0\mathbf{P}_{3d}$. 

Solving the generalized Anderson impurity problem is a difficult task, and at present there is
no universal impurity solver that works efficiently and accurately in all parameter regimes.
Here we make use of impurity solvers based on an expansion in the hybridization strength given 
by $\Delta_{3d}(\omega)$ around the atomic limit. The starting point is an exact diagonalization
of the (isolated) impurity subspace i.e. the Co $3d$-shell in our case given by the interacting
Hamiltonian $\hat{\mathcal H}_{3d}$. The hybridization of the impurity subspace with the rest of 
the system (given by the hybridization function $\Delta_{3d}(\omega)$) is then treated perturbatively. 

The so-called Non-Crossing Approximation (NCA) is a self-consistent perturbation expansion to lowest 
order in the hybridization strength. NCA only takes into account the most simple diagrams describing 
simple hopping processes where an electron or hole hops into the impurity at some time and then out 
at a later time (see Fig. \ref{fig:nca+oca} in App. \ref{app:nca+oca}). 
The so-called One-Crossing Approximation (OCA) improves upon the NCA by taking into account 
second order diagrams where two additional electrons (holes) are acommodated on the impurity 
at the same time in addition to the NCA diagrams as shown in Fig. \ref{fig:nca+oca}. 
OCA is the lowest order self-consistent approximation that is exact up to first order in the 
hybridization $\Delta_{3d}\propto V^2$ for all physical quantities while NCA is only exact
to first order in $\Delta_{3d}$ for the conduction electron self-energy. 
OCA improves considerably many of the shortcomings of NCA: It substantially improves the width
of the Kondo peak and hence the Kondo temperature which now are only slightly underestimated.
It also corrects the asymmetry of the Kondo peak. For very low temperatures ($T \ll T_K$), however, 
the height of the Kondo peak is still overestimated, and the Fermi liquid behaviour at zero 
temperature is not recovered. 

Hence, OCA is a reasonable approximation for solving the generalized impurity problem
as long as the temperatures are not too low (i.e. more than one order of magnitude below $T_K$).
In App. \ref{app:nca+oca} we give a brief introduction to the NCA and OCA impurity solvers.
A detailed description of the NCA and OCA methods can be found e.g. in Refs.
\onlinecite{Kotliar:rmp:2006,Haule:2009,Hewson:book,Haule:prb:2001}.

%%%%%%%%%%%%%%%%%%%%%%%%%%%%%%%%%%%%%%%%%%%%%%%%%%%%%%%%%%
\section{DFT atomic and electronic structure calculations}
\label{sec:dft-results}
%%%%%%%%%%%%%%%%%%%%%%%%%%%%%%%%%%%%%%%%%%%%%%%%%%%%%%%%%%

First, we have calculated the adsorption curves for a Co atom on a 
Graphene sheet with density-functional calculations 
within the generalized gradient approximation (GGA) with the CRYSTAL06
\cite{CRYSTAL06} {\it ab initio} electronic structure program for periodic 
systems employing an elaborate all-electron Gaussian basis set (6-31g).

\begin{figure}
  \begin{tabular}{c}
    \includegraphics[width=0.8\linewidth]{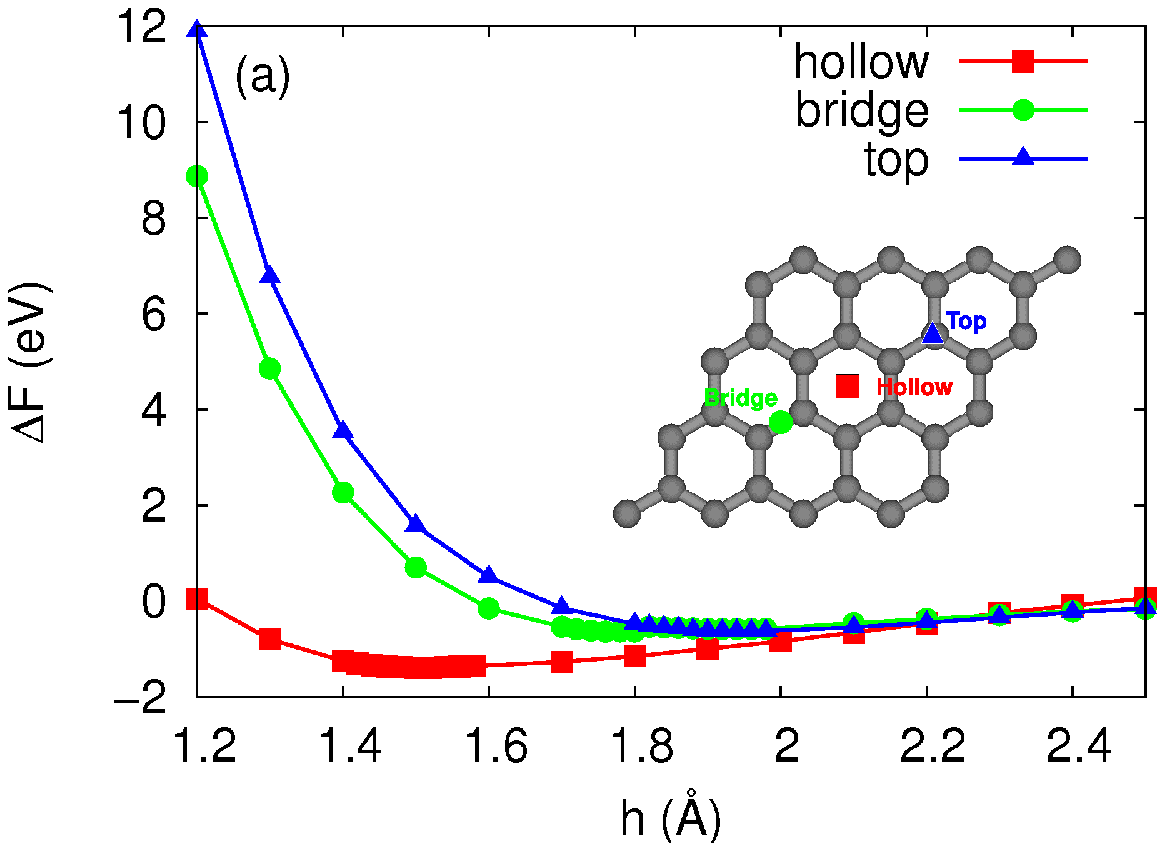} \\
    \includegraphics[width=0.8\linewidth]{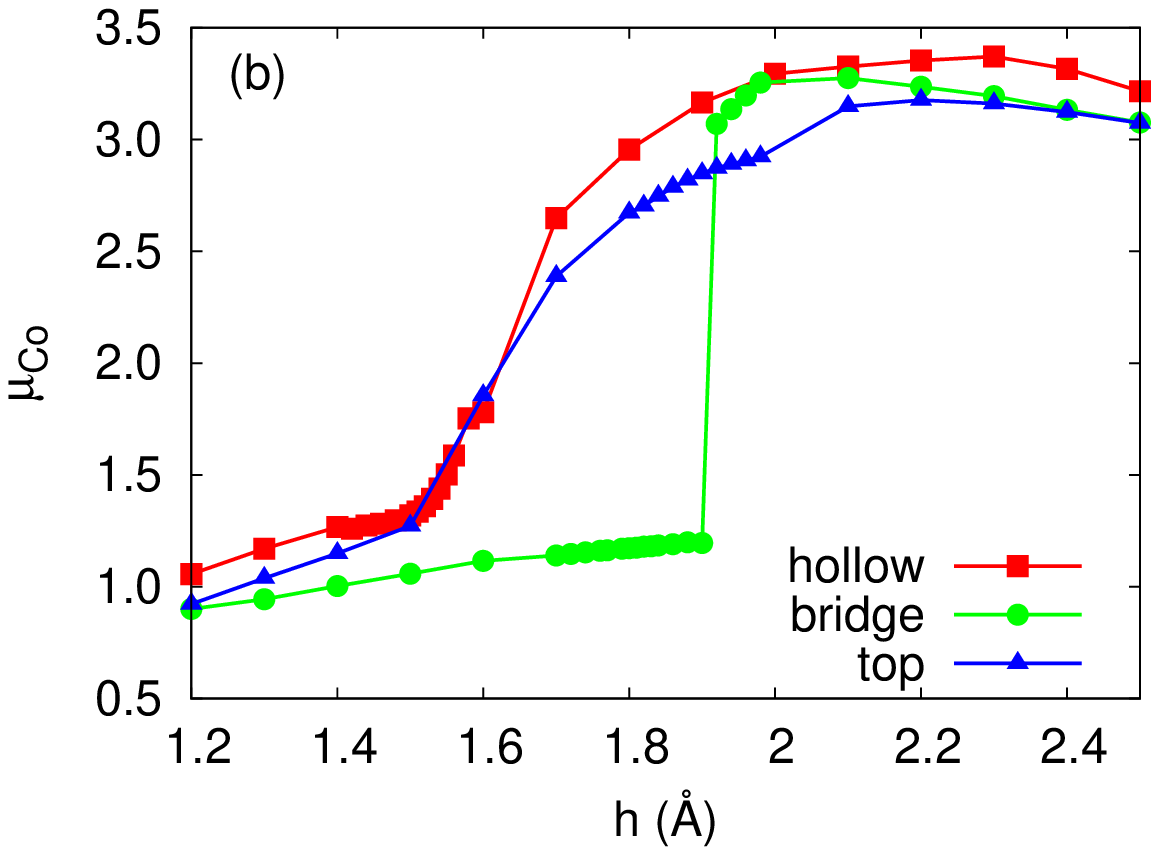} 
  \end{tabular}
  \caption{(color online)
    Adsorption of Co atoms on graphene. (a) Adsorption energy of 
    Co adatom on graphene sheet in dependence of distance $h$ for 
    the three different adsorption sites as shown in the inset:
    Hollow (h), top (t), and bridge site (b).
    (b) Magnetic moment of Co adatom on graphene sheet in 
    dependence of distance $h$ between Co atom and graphene 
    sheet.
  } 
  \label{fig:co-adsorption}
\end{figure}

As possible adsorption sites for the Co impurity we consider the three 
high symmetry sites shown on the inset of the top panel of Fig. 
\ref{fig:co-adsorption}, i.e. the hollow, top, and bridge site. For the 
calculations we have chosen a 5x5 unit cell in order to avoid interactions
between impurities in different unit cells. We have checked that the
results essentially do not change when choosing a slightly smaller (4x4) 
or bigger (6x6) unit cell.
Fig. \ref{fig:co-adsorption}(a) shows the adsorption energy 
for a Co atom in dependence of the distance $h$ between the Co atom from 
the graphene sheet for the three different sites. One can see that the 
hollow site is the most favourable absorption site. We find the energy 
minimum for this site at a distance of $h=$1.51\r{A}. As can be seen the 
energy minima for the other two adsorption sites are much more shallow. 
We find $h=1.81$\r{A} for the bridge site, and $h=1.96$\r{A} for the top site.
The top site is somewhat more favourable in energy than the bridge site.

\begin{table}
  \begin{tabular}{l|c|c|c|c|c|c|c}
    & h (\r{A}) & $E_{\rm ad}$ (eV) & $\Delta Q_{\rm Co}$ ($e$) & $\mu_{\rm Co}$ & $\mu$ & $N_{3d}$ & $\mu_{3d}$ \\
    \hline
    h        & 1.51 & -1.395 & 0.536 & 1.336 & 1.127 & 7.530 & 1.197 \\
    b        & 1.81 & -0.660 & 0.354 & 1.175 & 1.109 & 7.533 & 1.627 \\
    t        & 1.96 & -0.628 & 0.248 & 2.905 & 2.913 & 7.365 & 2.150 \\
    h$^\ast$ & 1.87 & ------ & 0.311 & 3.065 & 2.890 & 7.337 & 2.150 \\
  \end{tabular}
  \caption{
    Summary of GGA results. Optimal height ($h$) of Co atom, adsorption energy 
    ($E_{\rm ad}$), charge ($\Delta Q_{\rm Co}$) and magnetic moment ($\mu_{\rm Co}$) 
    of Co atom, total magnetic moment of unit cell ($\mu$), occupation ($N_{3d}$) 
    and magnetic moment $\mu_{3d}$ of Co $3d$-shell for the different
    adsorption sites considered here, i.e. h (hollow), b (bridge), t (top).
    In the last row (h$^\ast$), we show the GGA results for the Co atom at the 
    hollow site for the optimal distance $h=1.87$\r{A} calculated with the
    B3LYP hybrid functional (see text).
  }
  \label{tab:gga-results}
\end{table}

In Fig. \ref{fig:co-adsorption}(b) the magnetic moment of the Co atom
in dependence of the height $h$ of the Co atom over the graphene sheet 
is shown. Initially the magnetic moment is 3, i.e. that of a free Co 
atom. When the Co atom approaches the equilibrium distance for a certain 
adsorption site the magnetic moment starts to decrease. 
At the equilibrium distance, the magnetic moment of the Co atom has decreased to 
below 1.5 for the hollow site while for the bridge site it is more close to
1 already. For the top site , on the other hand, the  magnetic moment is still 
around 3 at the equilibrium distance as for the free atom. 
The magnetic moment for all three adsorption sites approaches 1 when 
further decreasing the height of the Co atom.
In the case of the bridge adsorption site the transition from the 
free magnetic moment of 3~$\mu_B$  to a magnetic moment of 1~$\mu_B$
is remarkably abrupt. A summary of our GGA results is given in Tab. 
\ref{tab:gga-results}. Our GGA results are in good agreement with 
previous work by other groups using different methods 
\cite{Wehling:prb:2010a,Mao:jphyscm:2008}.
In particular, we find for the hollow adsorption site at the equilibrium distance 
that the magnetic moment of the $3d$-shell is indeed close enough to 1, implying a 
spin-1/2 for the Co $3d$-shell, as reported by Wehling {\it et al.} \cite{Wehling:prb:2010a}. 
However, the occupation of the $3d$-shell is around 7.5 and not around 9 as reported by
Wehling {\it et al.} 
As will be explained below the approximate spin-1/2 of the Co $3d$ is essentially 
due to a hole in the E$_1$ minority-spin levels.

We have also performed hybrid functional calculations\cite{Becke:jcp:1993} which 
in general yield better geometries for molecules. Using the GAUSSIAN09\cite{G09} 
quantum chemistry code we have opimized the distance $h$ of the Co atom on the 
hollow site of a finite graphene flake of hexagonal shape with 24 C atoms and with 
H atoms at the borders to saturate the bonds \cite{footnote1}.
Employing the popular B3LYP hybrid functional together with the 6-31G 
basis set (also employed before in the CRYSTAL06 calculations) we now 
find a considerably higher value for the optimal height of the Co atom 
at the hollow site, namely $h=1.87$\r{A}.
For the GGA functionial we obtain very similar results ($h=1.47$\r{A})
as before with CRYSTAL06 code.
At the optimal height calculated with B3LYP ($h=1.87$\r{A}) both, the GGA and the B3LYP 
results are remarkably similar with respect to the magnetic moments and occupations of 
the Co $3d$-shell: The occupation of the $3d$-shell is 7.4 and the corresponding magnetic 
moment is close to 2 now, implying a total spin of 1 for the $3d$-shell.
In the last row of Tab. \ref{tab:gga-results}, we have added the GGA results for the 
hollow site with the Co atom at the optimal distance $h=1.87$\r{A} calculated with the 
B3LYP functional.

\begin{figure}
  \begin{center}
    \includegraphics[width=0.9\linewidth]{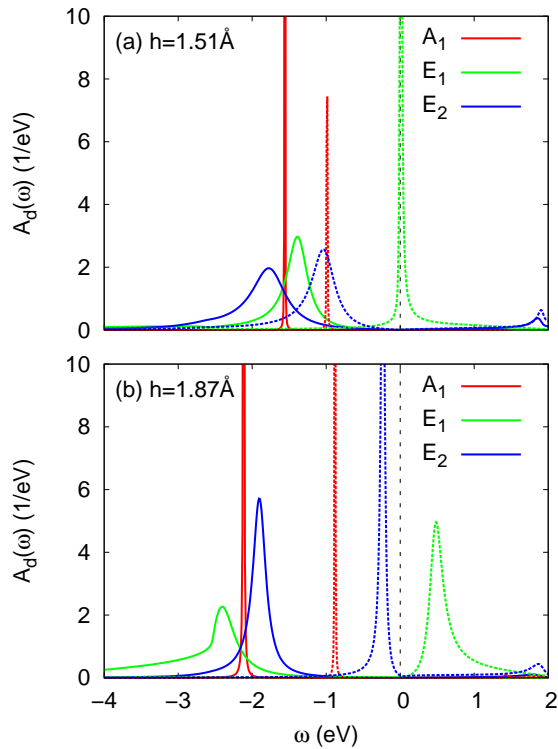}
  \end{center}
  \caption{(color online)
    Co $3d$-spectral densities calculated with 
    spin-polarized GGA for Co atom at hollow site
    for (a) the GGA equilibrium distance $h=1.51$~\r{A}
    and (b) the B3LYP equlibrium distance $h^\ast=1.87$~\r{A}. 
    Full (dashed) lines correspond to spin-up (down) states. 
    The vertical dashed line indicates the Fermi level.
  }
  \label{fig:gga-dos}
\end{figure}

In Fig. \ref{fig:gga-dos}, we show the Co $3d$-spectral functions 
calculated with spin-polarized GGA for the Co atom at the hollow site 
for two different heights, i.e. the optimal height predicted by GGA 
($h=1.51$~\r{A}) and the optimal height ($h=1.87$~\r{A}) predicted 
by B3LYP. 
Due to the C6-symmetry of the system the Co $3d$-shell splits into three 
symmetry groups: The non-degenerate A$_1$ group consisting of the rotationally 
invariant $3d_{3z^2-r^2}$-orbital only, the doubly degenerate E$_1$ group consisting 
of the $3d_{xz}$- and $3d_{yz}$-orbitals and the E$_2$ group consisting of the 
$3d_{xy}$- and $3d_{x^2-y^2}$-orbitals. 
The postitions of the peaks in the spectral functions show the effective
energy levels for each of the symmetry groups.
We can see that at the GGA equilibrium height (Fig. \ref{fig:gga-dos}(a)) the spin-1/2 
is essentially due to a hole in the  E$_1$-levels. The E$_1$ minority-spin levels
are exactly at the Fermi level and hence are half-filled, while the E$_1$ 
majority-spin and the A$_1$ and E$_2$ minority- and majority-spin levels are well 
below the Fermi level. This level scheme seems to suggest a $3d^9$ configuration
as reported by Wehling {\it et al.} \cite{Wehling:prb:2010a}, although the
true occupation of the $3d$-shell is really only 7.5. The reason for this 
apparant discrepancy is the {\it dynamic} hybridization of the Co $3d$-levels with 
the graphene substrate.

The Co $3d$-spectral density calculated at the B3LYP equilibrium distance
is shown in Fig. \ref{fig:gga-dos}(b). The main difference with the spectral
density at the GGA equilibrium (Fig. \ref{fig:gga-dos}(a)) is that the peak
corresponding to the minority-spin E$_1$-levels are now completely above the
Fermi level, and hence the minority-spin E$_1$-levels are completely empty.
Hence the $3d$-shell now yields a magnetic moment of about 2, i.e. a spin-1
due to two holes in the E$_1$-levels. 
Note that now the total magnetic moment of the Co atom is 3 due to a single 
electron in the Co $4s$-orbital which is fully spin-polarized. 
Also note that in this geometry, both the A$_1$- and the $4s$-level do not 
couple at all to the graphene substrate for symmetry reasons 
\cite{Wehling:prb:2010a}. 
Our GGA results at the B3LYP equilibrium distance show some similarity to the 
GGA+U results of Wehling {\it et al.}\cite{Wehling:prb:2010b}

%%%%%%%%%%%%%%%%%%%%%%%%%%%%%%%%%%%%%%%%%%%%%%%%%%%%%%%%%%%%%%%%%%
\section{GGA+OCA calculations for Co atom on hollow site}
\label{sec:oca-results}
%%%%%%%%%%%%%%%%%%%%%%%%%%%%%%%%%%%%%%%%%%%%%%%%%%%%%%%%%%%%%%%%%%

\begin{figure}
  \begin{center}
    \includegraphics[width=0.9\linewidth]{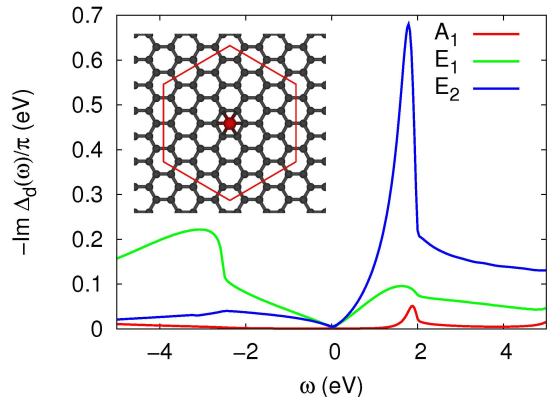}
  \end{center}
  \caption{(color online)
    Imaginary part of the hybridization functions
    for the Co atom at the hollow site as shown in the
    inset for $h=1.87$~\r{A} and for the three symmetry 
    groups.
    The inset shows a graphene sheet with Co adatom at hollow site 
    The red hexagon indicates the central region C containing
    54 carbon atoms. 
  }
  \label{fig:graphene+co}
\end{figure}

The inset of Fig. \ref{fig:graphene+co} shows the central region 
C containing the Co adatom at the hollow site embeded into a perfect 
graphene sheet. The cell has been taken large enough in order to avoid 
interaction between Co atoms in the CRYSTAL periodic supercell calculation.
The shape of the unit cell has been chosen to reflect the 6-fold symmetry 
of the system. The C region can now be embeded directly into the graphene
lattice set up by supercells of the same shape as C as described before 
in Sec. \ref{sec:method}.

\begin{table}
  \begin{tabular}{c|c|c|c}
    & $\epsilon_{3d}^{\rm KS}$ (eV) & ${\rm Re}\;\Delta_{3d}$ (eV) & $\tilde\epsilon_{3d}^{\rm KS}$ (eV) \\
    \hline
    A$_1$ & -3.650 & -0.144 & -3.794 \\
    E$_1$ & -3.637 & +0.197 & -3.440 \\
    E$_2$ & -3.414 & -0.389 & -3.803
  \end{tabular}
  \caption{
    Crystal field splittings for $3d$-shell of Co atom on hollow site of graphene ($h=1.87$~\r{A}).
    $\epsilon_{3d}^{\rm KS}$ are the on-site energies before double-counting correction,
    ${\rm Re}\;\Delta_{3d}$ is the real part of the hybridization function at the Fermi 
    level, and $\tilde\epsilon_{3d}^{\rm KS}\equiv\epsilon_{3d}^{\rm KS}+{\rm Re}\Delta_{3d}(0)$
    are the effective on-site energies taking into account the real part of the 
    hybridization function. 
    Note that the A$_1$ and E$_2$ effective energy levels are almost degenerate.
  }
  \label{tab:crystal-fields}
\end{table}

Due to the C6-symmetry of the system the Co $3d$-shell splits into three 
symmetry groups: The non-degenerate A$_1$ group consisting of the rotationally 
invariant $3d_{3z^2-r^2}$-orbital only, the doubly degenerate E$_1$ group consisting 
of the $3d_{xz}$- and $3d_{yz}$-orbitals and the E$_2$ group consisting of the 
$3d_{xy}$- and $3d_{x^2-y^2}$-orbitals. 

The right panel of Fig. \ref{fig:graphene+co} shows the hybridization functions for 
each symmetry group. As can be seen at low energies the $3d_{3z^2-r^2}$-orbital (A$_1$) 
has an essentially flat and zero hybridization with the graphene sheet. This can be 
understood by the fact that the coupling of the $3d_{3z^2-r^2}$ on the hollow site 
to the $2p_z$-orbitals of the graphene sheet is zero due to symmetry considerations 
\cite{Wehling:prb:2010a}.
The E$_2$-group has the (overall) strongest coupling to the C $2p_z$-orbitals in that geometry 
as can be seen from the corresponding hybridization function. 
Note that the hybridization 
function for both the E$_1$ and E$_2$ symmetry group vanish linearly near the Fermi energy 
reflecting the linear DOS of the graphene sheet. 
Interestingly, the slopes for the E$_1$- and E$_2$-hybridization functions are different 
for energies below and above the Fermi level. Morover, for negative energies the 
E$_1$-hybridization function has a steeper slope than the E$_2$-hybridization, while for 
positive energies it is the other way round.  This implies that E$_1$ couples predominantly 
to the negative energy band of graphene while $E_2$ couples predominantly to the positive 
energy band.

In Table. \ref{tab:crystal-fields}, we show the energy levels, the real part
of the hybridization function at the Fermi level, and the effective energy 
levels of the Co $3d$-shell. While the splitting of the bare energy levels
$\epsilon_{3d}$ is around 0.2~eV, the splitting of the effective energy levels
$\tilde\epsilon_{3d}$ taking into account the (real part of the) hybridization 
with the graphene is somewhat bigger with around 0.4~eV. Taking into account 
the real part of the hybridization changes also the ordering of the energy levels.
For the ordering of the effective energy levels we obtain: 
$\tilde\epsilon(E_2)<\tilde\epsilon(A_1)<\tilde\epsilon(E_1)$. 
Note that this level ordering is similar to the one obtained by Wehling 
{\it et al.}\cite{Wehling:prb:2010b}.

%%%%%%%%%%%%%%%%%%%%%%%%%%%%%%%%%%%%%%%%%%
\subsection{GGA+OCA results}
%%%%%%%%%%%%%%%%%%%%%%%%%%%%%%%%%%%%%%%%%%

\begin{figure*}
    \includegraphics[width=\linewidth]{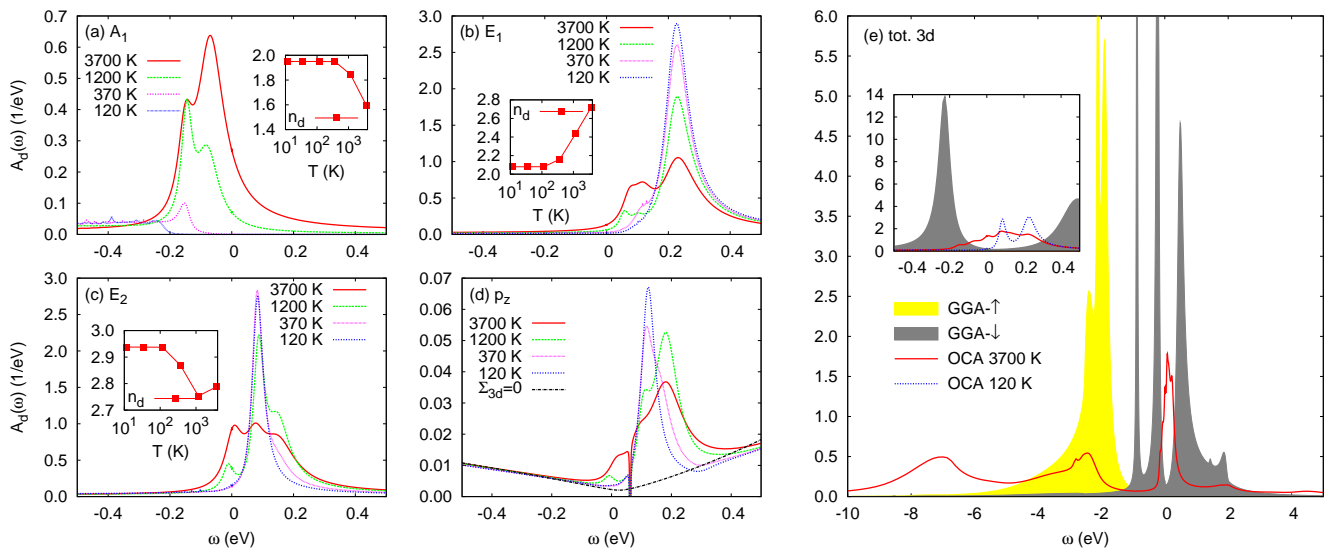}
    \caption{(color online)
      GGA+OCA results for $U=5$~eV and $J=0.9$~eV and total occupation $N_{3d}$ 
      of the $3d$-shell around 7.0.
      (a-c) Co $3d$-spectral functions near the Fermi level for all three symmetry groups
      at different temperatures. The insets show the corresponding orbital occupations 
      as a function of temperature. 
      (d) Spectral function of $p_z$-orbital of carbon atoms near the Co-atom for different 
      temperatures. The black dashed line shows the $p_z$-spectral function when the self-energy 
      of the Co $3d$-electrons $\Sigma_{3d}$ is set to zero.
      (e) Comparison of GGA+OCA $3d$-spectral density with spin-polarized GGA spectral density
      on a large energy scale. The inset shows a blow-up of the region near the Fermi level.
    }
    \label{fig:oca-results}
\end{figure*}

We now solve the generalized impurity problem of the Co $3d$-shell coupled to the graphene
sheet by using the OCA impurity solver as described in Sec. \ref{sec:method} and in App. 
\ref{app:nca+oca}. We take the Coulomb 
repulsion $U$ between $d$-orbitals to be somewhat higher than in bulk due to the supposedly 
smaller screening: $U=5$~eV. The Hund's rule coupling is taken to be the same as in bulk, i.e.
$J=0.9$~eV since it is less affected by screening effects. The DCC was calculated by
the standard expression given by eq. (\ref{eq:hdcc}) and taking the GGA value of about 7.5 
for $N_{3d}$.
With OCA, the total occupation of the impurity $3d$-shell is now between 
7.1 and 7.0 at low temperatures compared to the GGA value of 7.5.
This result is very typical for correlated calculations: The correlations
push the system towards integer occupation numbers to lower the energy.

Fig. \ref{fig:oca-results} shows the spectral functions of the Co $3d$-electrons
for different temperatures calculated with OCA. At high temperatures we find
peaks near the Fermi level for all three symmetry groups. However, the temperature
dependencies of the individual peaks are quite different from each other:
For the A$_1$-symmetry consisting of the $d_{3z^2-r^2}$-orbital the peak becomes 
smaller at increasing temperatures and finally develops a step-like behaviour at 
low temperatures. The peaks in the E$_1$ and E$_2$-spectra on the other hand become 
more pronounced with decreasing temperature more or less as expected for Kondo effect. 

The form of the spectral functions and the temperature behaviour can be understood
by looking at the occupation numbers for the different symmetry groups. 
The insets of Figs. \ref{fig:oca-results}(a-c) show the occupation numbers as a function 
of temperature for each of the three symmetry groups.
For the A$_1$-symmetry the occupation changes from $\approx$1.6 at high temperatures to
almost 2 at low temperatures. Hence, at low temperatures the $d_{3z^2-r^2}$-orbital
is practically completely filled, and excitations near the Fermi level are suppressed.
For the E$_1$-symmetry group composed of the $d_{xz}$- and $d_{yz}$-orbitals the occupation
changes from 2.8 at high temperatures to 2.1 at low temperatures. 
Hence at low temperatures this doubly-degenerate channel is essentially half-filled,
corresponding to a spin of almost 1 in this channel due to Hund's rule coupling. 
The peak near the Fermi level in the E$_1$-spectral function is hence due to a spin-1
Kondo effect in this channel. 
For the E$_2$-symmetry composed of the $d_{xy}$- and $d_{x^2-y^2}$-orbitals we find an occupation
of 2.8 for high temperatures, and of almost 3 for low temperatures. 
Hence the doubly-degenerate E$_2$-channel accomodates 3 electrons with a total spin-1/2
so that the peak in the E$_2$-spectral function is due to a spin-1/2 Kondo effect.
Accordingly, the peaks near the Fermi level in the $3d$-spectral function 
at low temperature are due to a spin-3/2 Kondo effect in the $3d$-levels. 
This interpretation is further corroborated by the fact that the main contribution 
to the electronic configuration of the $3d$-shell is an eightfold degenerate state 
with $N_d=7$ and $S=3/2$.
The very fact that such a high spin state can be screened by the graphene substrate
is an evidence of the multi-channel character of the graphene as a conduction electron
bath.

In Fig. \ref{fig:oca-results}(e) we compare the $3d$-spectral density calculated with 
the GGA+OCA (at $T=120$~K) method on the one hand and with spin-polarized GGA on the other
hand. As can be seen the spectral densities calculated by the two methods differ 
significantly from each other. In the GGA spectrum, the peak just below the Fermi level
(see also the inset of Fig. \ref{fig:oca-results}(e)) corresponds to a completely filled 
$E_1$-shell with down-spin as can be seen by comparison with Fig. \ref{fig:graphene+co}(c).
The peak above (and also a little farther from) the Fermi level corresponds to a completely 
empty spin-down $E_2$-shell. Both peaks carry essentially the full spectral weight of the 
correpsonding orbitals. In contrast, the two peaks near the Fermi level in the GGA+OCA 
spectra which are also of $E_1$- and $E_2$-symmetry, do not carry the full spectral 
weight of the corresponding orbitals due to the renormalization of the quasi-particle
by the electron-electron interactions which cannot be captured by a static mean-field
calculation like GGA.

Fig. \ref{fig:oca-results}(d) shows that the correlations in the Co $3d$-shell also 
affect the projected DOS of the $p_z$-orbitals on the carbon atoms that are directly 
coupled to the Co atom (i.e. the 6 carbon atoms nearest to the Co atom). This can be
understood as follows: Due to the coupling between the carbon $p_z$- and the Co 
$3d$-orbitals, the full $p_Z$-Green's function $G_p(\omega)$ is related to the full 
Green's of the $3d$-orbitals $G_d(\omega)$ by:
\begin{equation}
  \label{eq:}
  G_p(\omega) = G_p^0(\omega) + \sum_d G_p^0(\omega) V_{pd} G_d(\omega) V_{dp} G_p^0(\omega)
\end{equation}
where $G_p^0(\omega)$ is the Green's function of the (unperturbed) graphene $p_z$-orbitals without
the coupling to the Co $3d$-orbitals and $V_{pd}=V_{dp}^\ast$ is the coupling between the $p_z$- and
individual $3d$-orbitals. The full $3d$-Green's function $G_d(\omega)$ takes into account the 
correlations given by $\Sigma_{3d}(\omega)$ {\it and} the hybridization with the $p_z$-orbitals
given by $\Delta_{3d}(\omega)$, i.e. $V_{pd} G_d(\omega) V_{dp}$ represents the full $T$-matrix for the 
$p_z$-electrons.

One can see a temperature-dependent peak roughly at the position
of the peaks in the E$_1$- and E$_2$-spectral density. For comparison we also show
the $p_z$-DOS when the self-energy of the Co $3d$-electrons $\Sigma_{3d}$ is set to zero.
In this case the $3d$-levels are far below the Fermi level and hence cannot affect the
$p_z$-DOS near the Fermi level. Hence the sharp resonance in the $p_z$-DOS vanishes.
This shows that the Kondo peaks in the Co $3d$-spectra can also be detected indirectly 
by spectroscopy of the nearby carbon atoms.

%%%%%%%%%%%%%%%%%%%%%%%%%%%%%%%%%%%%%%%%%%%%%%%%%%
\subsection{Dependence on $3d$-energy levels}
%%%%%%%%%%%%%%%%%%%%%%%%%%%%%%%%%%%%%%%%%%%%%%%%%%

\begin{figure}
  \includegraphics[width=\linewidth]{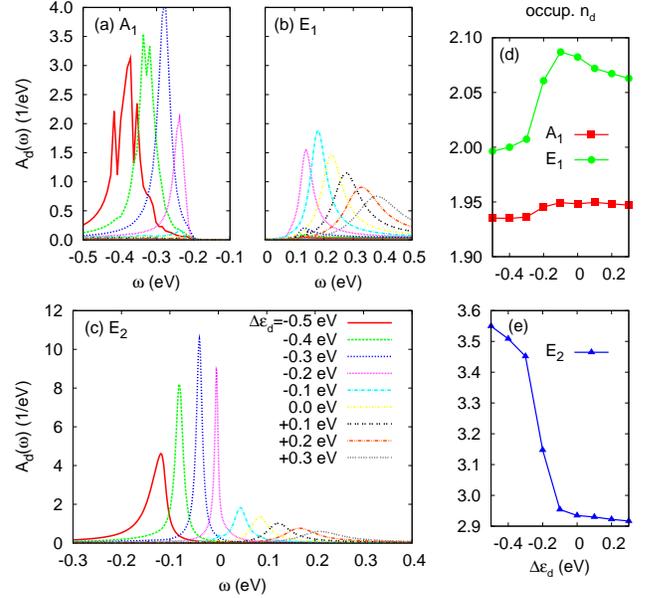} 
  \caption{(color online)
    $3d$-energy level dependence. (a-c) Spectral functions
    of Co $3d$-levels for different energy shifts $\Delta\epsilon_d$ 
    of the $3d$-levels (as indicated in the legend of (c)) 
    at $T=37$~K and for the three different symmetries.
    (d,e) Orbital occupancies of the Co $3d$-levels
    as a function of the energy shift for each of the three
    symmetries and for $T=37$~K.
  }
  \label{fig:ed-dependence}
\end{figure}

The double-counting correction term $\mathbf{h}_{\rm dcc}$
given by eq. (\ref{eq:hdcc}) which substracts out the Coulomb 
interaction within the correlated $3d$-subspace that has already been taken 
into account on a static mean-field level in the LDA or GGA calculation is not 
known exactly.  Hence the exact position of the $3d$-energy levels is not known. 
Here we explore how shifting the $3d$-energy levels changes the results. 

Figs. \ref{fig:ed-dependence}(a-c) show the Co $3d$-spectral densities for each
of the different symmetries and for different shifts $\Delta\epsilon_d$ of the
$3d$-levels with respect to their original position, and Figs. 
\ref{fig:ed-dependence}d,e the corresponding occupancies.
Quite obviously there is a drastic change in the correlations when the impurity-levels
are lowered by more than 0.2~eV. At that point, a substantial peak appears in the A$_1$-spectra,
while the peak in the E$_1$-spectra disappears. The peak in the E$_2$-spectra on
the other hand becomes qualitatively different for energy shifts $\Delta\epsilon_d\le-0.2$~eV,
i.e. their spectral weight increases considerably. The changes in the spectra are accompanied
by changes in the occupancies: Most importantly, the occupancy of the E$_2$-levels increases
abruptly by about 0.6 for $\Delta\epsilon_d\approx-0.2$~eV. The changes in the $3d$-levels
with A$_1$- and E$_1$-symmetry are less pronounced but also occur at the same point.
Also the overall occupation of the impurity levels changes quite abruptly by about 
0.5 electrons from $N_d\approx7$ to $N_d\approx7.5$. This indicates a strong change
in the electronic configuration of the Co $3d$-shell.

Indeed we find that for an energy shift around -0.2~eV, another $3d$-shell state
with $N_d=8$ and $S=1$ becomes similar in energy to the state with $N_d=7$ and 
$S=3/2$. Hence, at this point the system enters the mixed-valence regime. 
This explains the strong increase of the 
spectral weight in the E$_2$-peak for $\Delta\epsilon_d\le-0.2$~eV since the mixed-valence 
regime is characterized by a strong peak near the Fermi level. 
On the other hand, the integer occupation of 2 in the E$_1$-levels, and the associated
spin-1, essentially does not change upon entering the mixed-valence regime. 
In fact, the occupation is now even closer to 2 than it was in the Kondo regime. 
Also the occupation of the A$_1$-level remains essentially constant upon entering
the mixed-valence regime. Hence the mixed-valence is actually only induced in the 
E$_2$-levels since the occupations of the A$_1$-level and the E$_1$-levels in the
$N_d=8$ state are the same as in the $N_d=7$ state.
The vanishing of the peak in the E$_1$-spectra upon entering the mixed-valence 
regime is due to the weaker hybridization with the graphene for negative energies 
(compare Fig. \ref{fig:graphene+co}): As the Kondo peak in the E$_1$-levels moves beyond 
the Dirac point as the impurity-levels are lowered in energy, the Kondo coupling
in the E$_1$-levels becomes weaker and the Kondo temperature much smaller so 
that the Kondo effect is not observed anymore at the temperatures considered here.
It is not quite clear what triggers the peak in the A$_1$-spectra upon entering the 
mixed-valence regime. One possibility is that the increased charge fluctuations in 
the mixed valence regime also lead to charge fluctuations in this level via the 
Hund's rule coupling to the other $3d$-levels. The slight decrease in the occupation
of the A$_1$-level in the mixed-valence regime supports this view.

Note that the position of a peak (if present) changes according to the energy shift.
This also explains the change in the width of the peaks for the E$_1$- and E$_2$-
symmetries: As a peak moves away from the Fermi level the broadening increases due 
to the increased hybridization with the graphene substrate. The A$_1$-level on
the other hand does not couple at all to the graphene substrate (zero hybridization)
so that the changes in the shape of the A$_1$-peak are purely correlation effects.

Hence tuning the energies of the Co $3d$-levels changes the correlations in these
levels due to changes in the occupancy. More precisely, the system is driven from
the Kondo regime to the mixed-valence regime when the energies of the $3d$-levels 
are lowered by more than 0.2~eV with respect to their original energies.
The change in the correlations is reflected by corresponding changes in the 
$3d$-spectral functions.

%%%%%%%%%%%%%%%%%%%%%%%%%%%%%%%%%%%%
\subsection{Gate voltage dependence}
%%%%%%%%%%%%%%%%%%%%%%%%%%%%%%%%%%%%

\begin{figure*}
  \includegraphics[width=\linewidth]{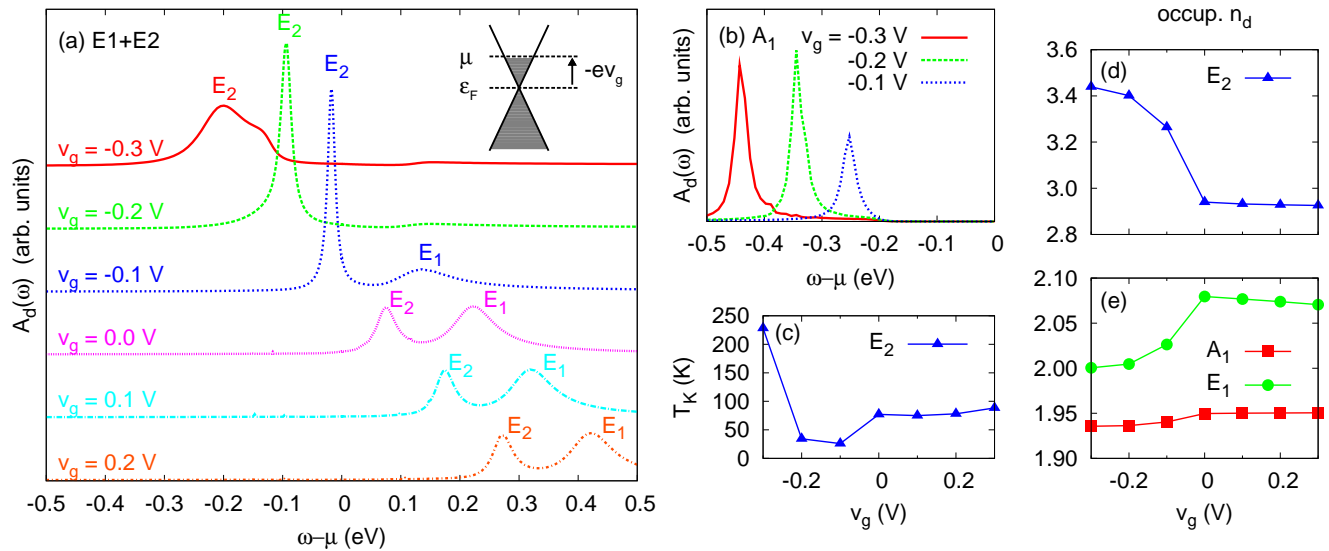} 
  \caption{(color online)
    Gate voltage dependence. 
    (a) $3d$-spectral function for E$_1$ and E$_2$ symmetry 
    for different gate voltages $v_g$ ($T=37$~K). 
    Different offsets have been added to the spectral functions of different $v_g$ 
    in order to better distinguish them from each other.
    The inset shows a Dirac cone of the graphene sheet 
    with the position of the chemical potential $\mu$ 
    for a certain gate voltage $v_g$.
    (b) Spectral function for Co $3d$-level of A$_1$-symmetry
    for different gate voltages $v_g$ ($T=37$~K). 
    The chemical potential in both (a) and (b) is always at zero frequency.
    (c) Kondo temperature estimated from width of E$_2$ resonance
    as a function of gate voltage $v_g$ at $T=37$~K. 
    (d-e) Occupancies of Co $3d$-levels resolved for the
    three different symmetries as a function of gate voltage $v_g$
    at $T=37$~K. 
  }
  \label{fig:gate-voltage}
\end{figure*}

In the previous subsection we have seen that changing the exact position
of the Co $3d$-levels with respect to the Fermi level has a considerable 
effect on the electronic correlations and hence on the spectral density 
of the $3d$-levels. Although changing the impurity levels directly is not 
feasible in an experiment, one can instead easily tune the chemical potential 
of graphene by application of a gate voltage. The chemical potential then is 
shifted with respect to both, the impurity levels and the graphene bands.
Hence it should be possible to control the electronic correlations and in 
particular the Kondo effect by application of a gate voltage. The change in 
correlations can then be detected in the $3d$-spectra for example by STM 
spectroscopy.

More precisely, a gate voltage $v_g$ shifts the chemical potential $\mu$ 
with respect to the graphene bands and the impurity levels according to 
$\mu=\epsilon_F-e v_g$ where $e$ denotes the elementary charge and $\epsilon_F$ 
is the Fermi level of neutral graphene, i.e. the energy of the Dirac points 
(see inset of Fig. \ref{fig:gate-voltage}a). Overall the results for shifting 
the chemical potential presented in Fig. \ref{fig:gate-voltage} look very similar 
to the previous ones for shifting the impurity levels in Fig. \ref{fig:ed-dependence}. 

At the hollow site only the Co $3d$-levels of E$_1$- and 
E$_2$-symmetry couple to the graphene sheet for low energies while the A$_1$-level
is completely decoupled. Hence only the spectra of E$_1$- and E$_2$-levels can
be measured by an STM. Therefore in Fig. \ref{fig:gate-voltage}(a), we show the
total spectral density of the Co E$_1$- and E$_2$-levels for different values of the 
gate voltage. 
For gate voltages smaller than -0.1~eV there is only one
peak in the spectral function which is due to the E$_2$-levels.
For gate voltages $\ge-0.1$~eV a second peak appears which originates
from the E$_1$-levels. 
The E$_2$-peak and also the E$_1$-peak (when present) move with respect to the 
chemical potential $\mu$ according to the change in gate 
voltage. Hence the peaks in the spectral function are not pinned to the
chemical potential as expected for normal Kondo effect, but are rather pinned to
the Dirac points of the graphene sheet. 
For completeness we also show the spectral density of the A$_1$-level in Fig.
\ref{fig:gate-voltage}b for different gate voltages. A peak in the spectral
density at low temperature is only observed for negative bias voltages. The 
peak disappears for zero and positive gate voltage. Also the resonance in the 
A$_1$-level appears to be pinned to the Dirac point of graphene rather than
to the chemical potential. 

As can be seen in Figs. \ref{fig:gate-voltage}d,e also the individual occupations 
of the $3d$-levels as a function of the applied gate voltage show a similar 
behaviour as before as a function of the energy shift $\Delta\epsilon_d$ of the 
$3d$-levels. In fact, the explanation for the behaviour of the spectra and occupancies
as a function of the energy shift of the $3d$-levels also applies to their behaviour
as a function of the gate voltage. Here, when the gate voltage becomes smaller
than -0.1~V the system leaves the Kondo regime with an $N_d=7$ and $S=3/2$ state
as the principal contribution to the electronic configuration of the $3d$-shell, 
and enters the mixed-valence regime with an equal contribution coming from
an $N_d=8$ and $S=1$ state.

We can estimate a Kondo temperature from the width of the resonance in the
Kondo regime. Fig. \ref{fig:gate-voltage}c shows the Kondo temperature
of the E$_2$-levels estimated from the resonance width $\Gamma$ using the relation
$k_BT_K=\pi\cdot w\cdot\Gamma/8$ known from the Fermi liquid theory of the Anderson model
where $w$ is Wilson's number 0.4128 (see e.g. the book by Hewson\cite{Hewson:book}, chap. 5).
In Fig. \ref{fig:gate-voltage}c we have also included ``Kondo temperatures''
estimated from the width of the peaks for negative gate voltages where
the system is not in the Kondo regime anymore. Hence outside the Kondo regime 
it is only a measure for the width of the resonance near the chemical potential.

Note that in the Kondo regime (i.e. for $v_g\ge 0$) the width of the resonance 
and hence the Kondo temperature are essentially constant. This behaviour
is different from the one before where we only had shifted the impurity levels.
There the width of the E$_1$- and E$_2$-peaks increased as they moved away from 
the Fermi level due to the linear increase of the hybridization with the graphene 
substrate.
Here, however, the peak is pinned to the Dirac point of graphene, i.e. it moves 
together with the graphene bands with respect to the chemical potential. Hence the 
hybridization with the graphene conduction electrons does not change here.
The strong dependence of the width of the resonance on the gate voltage 
for negative gate voltages on the other hand is due to the changes in the 
electronic correlations as the system is driven out of the Kondo regime.

%%%%%%%%%%%%%%%%%%%%%%%%%%%%%%%%%%%%
\section{Summary and Conclusions}
\label{sec:conclusions}
%%%%%%%%%%%%%%%%%%%%%%%%%%%%%%%%%%%%

Using DFT electronic structure calculations 
we have found that a single Co atom most likely adsorbs 
at the hollow site of an otherwise clean graphene sheet. 
Furthermore the DFT calculations show, that the magnetic 
moment and the electronic configuration of the $3d$-shell 
of the Co atom depends quite strongly on the adsorption height 
over the graphene sheet. 
At large distantes the formation of a local moment is favoured
while at short distances the magnetic moment is quenched due
to strong hybridization with the substrate.
Since the adsorption energy curves are quite shallow, 
it might be possible in an experiment to push the Co atom 
from the local moment to the weakly correlated regime.

Using our GGA+OCA correlated electronic structure method 
for nanoscopic systems, we have then studied the impact 
of dynamical correlations due to the strongly interacting
Co $3d$-electrons on the electronic structure. Because of the 
strong crystal field splitting of the $3d$-levels by graphene
substrate, the behaviour of each of the Co $3d$-levels strongly 
depends on its orbital symmetry. 

We find that the dynamic correlations can give rise to 
a spin-3/2 Kondo effect. The spin-3/2 is composed of a 
spin-1 due to two electrons within the doubly-degenerate 
orbitals of E$_1$-symmetry and a spin-1/2 due to a hole 
within the doubly-degenerate orbitals of the E$_2$-symmetry.
The Kondo effect gives rise to corresponding peaks near the 
Fermi level in the E$_1$- and E$_2$-spectral function.
The A$_1$-channel on the other hand is nearly completely
filled and hence in the Kondo regime, it does not give rise 
to a peak near the Fermi level.
The realization of a spin-3/2 Kondo effect points to the 
multi-channel character of graphene. Hence a spin-1/2 
impurity could in principle give rise to multi-channel Kondo physics 
and Non-Fermi liquid behaviour. 

The Kondo effect can be controlled by a gate voltage which changes 
the chemical potential and thereby the filling of the $3d$-levels.
The change in the correlations can be observed by a stark change in 
the spectra as the system is driven out of the Kondo regime and into 
the mixed valence regime.
Another remarkable result is that the Kondo peaks appear to be pinned 
to the Dirac point of the graphene substrate rather than to the 
chemical potential when a gate voltage is applied. 
This last finding is actually in agreement with very recent STM spectroscopy 
measurements of Co atoms on graphene which find peaks in the dI/dV 
curves that move according to the applied gate voltage\cite{Brar:2010}.
But it has not yet been pointed out by other theoretical work considering 
the dependence of the Kondo effect in graphene on a gate-voltage
\cite{Sengupta:prb:2008,Vojta:2010}.

%%%%%%%%%%%%%%%%%%%%%%%%%%%%%%%%%%%%%%%%%%%%%%%%%%%%%%%%%%%%%%%%%
\section*{Acknowledgments}
We are grateful to Sasha Balatzky, Kostyantyn Kechedzhi and 
R\'egis Decker for useful discussions, and to Kristjan Haule 
for providing us with the OCA impurity solver. Part of this 
research was carried out at the KITP in the program 
"Towards Material Design with Correlated Electrons". 
GK was supported by NSF-DMR-0906943. 
%%%%%%%%%%%%%%%%%%%%%%%%%%%%%%%%%%%%%%%%%%%%%%%%%%%%%%%%%%%%%%%%%

\begin{appendix}

%%%%%%%%%%%%%%%%%%%%%%%%%%%%%%%%%%%%%%%%%%
\section{Details of the embedding method}
\label{app:embedding}
%%%%%%%%%%%%%%%%%%%%%%%%%%%%%%%%%%%%%%%%%%

Here we show how to compute the embedding self-energy $\mathbf\Sigma_{\rm H}$ 
given in eq. (\ref{eq:GC}) that describes the dynamic hybridization of the 
central region C with the rest of the system (i.e. the graphene sheet).
First, we obtain an effective mean-field description of the host material 
from a GGA calculation for a supercell of clean graphene corresponding to 
the C region. 
From the Kohn-Sham Hamiltonian $\mathbf{H}_{\rm H}(\vec{k})$ of the host 
material in reciprocal space we calculate the $k$-dependent Green's 
function:
\begin{equation}
  \mathbf{G}_{\rm H}(\vec{k},\omega) = (\omega+\mu - \mathbf{H}_{\rm H}(\vec{k}) + i\eta)^{-1}
\end{equation}
The local KS Hamiltonian of the supercell is obtained by summing up 
$\mathbf{H}_{\rm H}(\vec{k})$ over all k-points within the Brillouin zone:
\begin{equation}
  \mathbf{H}_{\rm H,0} = \sum_{\vec{k}} \mathbf{H}_{\rm H}(\vec{k})
\end{equation}
Similarly, the local Green's function of the supercell is given by
summing up $\mathbf{G}_{\rm H}(\vec{k},\omega)$ over all k-points within the Brillouin zone:
\begin{equation}
  \mathbf{G}_{\rm H,0}(\omega) = \sum_{\vec{k}} \mathbf{G}_{\rm H}(\vec{k},\omega)
\end{equation}
On the other hand, we can write the local Green's function also in terms of the 
local Hamiltonian $\mathbf{H}_{\rm H,0}$  of the supercell and the embedding
self-energy (which describes the dynamic hyrbidization with the rest of the
graphene) as: 
$\mathbf{G}_{\rm H,0}(\omega) = [\omega+\mu-\mathbf{H}_{\rm H,0}-\mathbf\Sigma_{\rm H}(\omega)]^{-1}$ .
Hence the embedding self-energy is given by:
\begin{equation}
  \mathbf\Sigma_{\rm H}(\omega) = \omega+\mu-\mathbf{H}_{\rm H,0}-\left[\mathbf{G}_{\rm H,0}(\omega)\right]^{-1}
\end{equation}
%%

%%%%%%%%%%%%%%%%%%%%%%%%%%%%%%%%%%%%%%%%%%
\section{The NCA and OCA impurity solvers}
\label{app:nca+oca}
%%%%%%%%%%%%%%%%%%%%%%%%%%%%%%%%%%%%%%%%%%

The general multi-orbital Anderson impurity model can be written in the
following form:
\begin{eqnarray}
  \label{eq:AIM}
  \hat{\mathcal{H}} &=& 
  \sum_{\alpha\beta} \epsilon^{d}_{\alpha}\,\hat{d}_{\alpha}^\dagger \hat{d}_{\alpha }
  + \frac{1}{2} \sum_{\alpha\beta\gamma\delta} U_{\alpha\beta\gamma\delta}
  \hat{d}_{\alpha}^\dagger \hat{d}_{\beta}^\dagger \hat{d}_{\gamma} \hat{d}_{\delta}
  \\
  &+& \sum_{k\nu\alpha\sigma} (V_{k\nu\alpha} \hat{c}_{k\nu}^\dagger \hat{d}_{\alpha}
  + V_{k\nu\alpha}^\ast \hat{d}_{\alpha}^\dagger \hat{c}_{k\nu})
  + \sum_{k\nu\sigma} \epsilon_{k\nu} \hat{c}_{k\nu}^\dagger \hat{c}_{k\nu}
  \nonumber
\end{eqnarray}
where in order to keep the notation simple we have combined the spin- and
orbital degrees of freedom into one index for each impurity level $\alpha$ 
and each band $\nu$. 

The Non-Crossing Approximation (NCA) and the One-Crossing 
Approximation (OCA) both solve the Anderson impurity model 
by expansion in the hybridization strength around the atomic 
limit. 
The starting point is an exact diagonalization of the impurity subspace,
i.e. of the Co $3d$-shell in our case, including the Hubbard-like 
interaction term:
\begin{eqnarray}
  \hat{\mathcal H}_{d} &\equiv& 
  \sum_{\alpha\beta} \epsilon^{d}_{\alpha}\,\hat{d}_{\alpha}^\dagger \hat{d}_{\alpha }
  + \frac{1}{2} \sum_{\alpha\beta\gamma\delta} U_{\alpha\beta\gamma\delta}
  \hat{d}_{\alpha}^\dagger \hat{d}_{\beta}^\dagger \hat{d}_{\gamma} \hat{d}_{\delta}
  \nonumber\\
  &\stackrel{\rm diag.}{\longrightarrow}& \sum_m \ket{m} E_m \bra{m} 
\end{eqnarray}
where $\ket{m}$ are the many-body eigenstates of $\hat{\mathcal H}_{3d}$ 
and $E_m$ the respective eigen energies. 

One now introduces auxiliary fields $\hat{a}_m$, $\hat{a}_m^\dagger$ (called pseudo-particles) 
such that each impurity state is represented by a corresponding pseudo-particle:
\begin{equation}
  \hat{a}_m^\dagger\ket{\rm PPV} \equiv \ket{m}
\end{equation}
where $\ket{\rm PPV}$ is the pseudo-particle vacuum. 
The completeness of the impurity eigenstates imposes the following constraint: 
\begin{equation}
  Q \equiv \sum_m \hat{a}_m^\dagger \hat{a}_m = 1
\end{equation}
The physical electron operators $\hat{d}_{\alpha}^\dagger$
can now be expressed by the PP operators:
\begin{equation}
  \hat{d}^\dagger_{\alpha} = \sum_{n,m} 
  %%\bra{n}\hat{d}^\dagger_{\alpha}\ket{m} 
  (F^{\alpha\dagger})_{nm}
  \hat{a}_n^\dagger \hat{a}_m
\end{equation}
where $(F^{\alpha\dagger})_{nm}\equiv\bra{n}\hat{d}^\dagger_{\alpha}\ket{m}$ are 
the matrix elements of the impurity-electron creation operator. 
For later convenience we also define the corresponding matrix elements
of the impurity-electron destruction operator as: 
$(F^{\alpha})_{nm}\equiv\bra{n}\hat{d}_{\alpha}\ket{m}$.
The anti-commutation rules for the physical electron operators then require that
the PP $\hat{a}_m$ is a boson (fermion) if the corresponding state $\ket{m}$ 
contains an even (odd) number of electrons.

In the PP representation we can now rewrite the Hamiltonian of the
generalized Anderson impurity model as follows:
\begin{eqnarray}
  \label{eq:PP-AIM}
  \hat{\mathcal H} &=& \sum_m E_m \hat{a}_m^\dagger \hat{a}_m 
  + \sum_{k\nu} \epsilon_{k\nu} \hat{c}_{k\nu}^\dagger\hat{c}_{k\nu}  +\lambda(Q-1)
  \nonumber\\
  &+& \sum_{{mn}\atop{k\nu\alpha}}
  \left(V_{k\nu,\alpha} \hat{c}_{k\nu}^\dagger \hat{a}_m^\dagger  (F^{\alpha})_{nm} \hat{a}_n + H.c. \right)  
  %%+V_{k\nu\alpha}^\ast \hat{a}_m^\dagger \hat{a}_n (F^{\alpha\dagger})_{nm}  \hat{c}_{k\nu}
\end{eqnarray}
where we have included the constraint $Q\equiv 1$ into the Hamiltonian.
The corresponding Lagrange multiplier $\lambda$ can be interpreted as a 
(negative) chemical potential for the PPs.

In the PP picture, the hybridization with the bath electrons
given by the last term in eq. (\ref{eq:PP-AIM}) becomes now the interaction 
for the PPs. Because of the fermionic and bosonic commutation rules for the 
PPs, one can now develop a diagrammatic perturbation expansion in the PP 
interaction. The PP propagators can then be written as
\begin{equation}
  G_m(\omega) = (\omega-\lambda-E_m-\Sigma_m(\omega))^{-1}
\end{equation}
where $\Sigma_m$ is the PP self-energy describing the dynamic interaction with
the other PPs.

\begin{figure}
  \includegraphics[width=\linewidth]{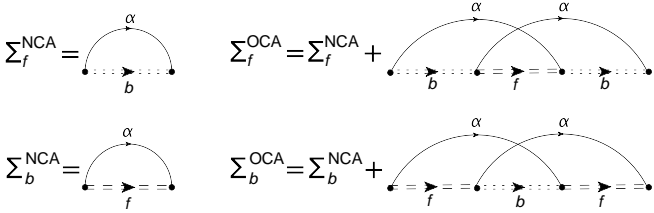}
  \caption{Diagrams for pseudo-particle self-energies in NCA and OCA approximations.}
  \label{fig:nca+oca}
\end{figure}

The Non-Crossing Approximation consists in taken into account the diagrams 
shown in the first row of Fig. \ref{fig:nca+oca} for a certain pseudo-particle.
The NCA diagrams describe processes where a single electron (hole) jumps from 
the bath to the impurity and back thereby temporarily creating a PP with N+1 
(N-1) electrons. The NCA equations correspond to a self-consistent perturbation 
expansion to lowest order in the hybridization function 
$\Delta_\alpha(\omega)\equiv\sum_{k,\nu}V_{k\nu,\alpha}^\ast  V_{k\nu,\alpha}$.
Since the fermionic self-energies depend on the dressed bosonic propagators, 
and vice versa, the NCA equations have to be solved self-consistently.
Once the NCA equations are solved the physical quantities can be calculated from the PP
self-energies. 

OCA takes into account second order diagrams where two bath electron lines cross 
as shown in the right column of Fig. \ref{fig:nca+oca}.
The self-energies for the fermions (bosons) again depend on the full propagators
for bosons (fermions), and hence the OCA equations have to be solved self-consistently.
The explicit expressions for the OCA self-energies are
second in the bath hybridization function $\Delta_\alpha$.
The OCA is the lowest order self-consistent approximation that is exact up to first order in the 
hybridization $\Delta_{3d}\propto V^2$ for all physical quantities while NCA is only exact
to first order in $\Delta_{3d}$ in the PP and conduction electron self-energies.
Further details of the NCA and OCA impurity solver can be found e.g. in Refs. 
\onlinecite{Haule:2009,Haule:prb:2001}.

\end{appendix}


\begin{thebibliography}{0}
\bibitem{footnote1} Unfortunately, we were not able to converge hybrid functional calculations with CRYSTAL06 for Co atoms on graphene sheets.
\end{thebibliography}
\end{document}